# An Outline of Security in Wireless Sensor Networks: Threats, Countermeasures and Implementations

M. Yasir Malik

*Institute of New Media and Communication*

*Seoul National University, Korea*

## ABSTRACT

With the expansion of wireless sensor networks, the need for securing the data flow through these networks is increasing. These sensor networks allow for easy-to-apply and flexible installations which have enabled them to be used for numerous applications. Due to these properties, they face distinct information security threats. Security of the data flowing through across networks provides the researchers with an interesting and intriguing potential for research. Design of these networks to ensure the protection of data faces the constraints of limited power and processing resources. We provide the basics of wireless sensor network security to help the researchers and engineers in better understanding of this applications field. In this chapter, we will provide the basics of information security with special emphasis on WSNs. The chapter will also give an overview of the information security requirements in these networks. Threats to the security of data in WSNs and some of their counter measures are also presented.

## 1. INTRODUCTION

Wireless sensor networks (WSNs) attract the attention of researchers and engineers thanks to their vast application scope. These allow for easy and flexible installation of wireless networks composed of large number of nodes. This gives WSN the capability to be used in unimaginable applications. They are finding their usages in habitat monitoring, manufacturing and logistics, environmental observation and forecast systems, military applications, health, home and office applications and a variety of intelligent and smart systems. Multimedia wireless sensor networking is a relatively new branch in this domain, which can process multimedia content i.e. still images, audio and video to name a few.

Such a sensor network is typically composed of hundreds, and sometimes thousands of nodes. These nodes are capable of receiving, processing and transmitting information, as based on the assigned tasks. Information flowing through WSN may be susceptible to eavesdropping, retransmit previous packets, injection of redundant or causeless bits in packets and many other threats of diverse nature. To ensure that the data being received and transmitted across these networks is secure and protected, information security plays a vital role.

As contrary to the Moore's law, there has been not much development in the hardware capacity and computational capabilities of the sensors being deployed in wireless sensor networks. These networks are kept inexpensive, thus introducing many constraints in the performance parameters. Low cost sensors incorporate shortcomings in their storage capacity, power requirements and processing speed.

This poses a unique dilemma for researchers as they have to design efficient and distinct information security schemes which work seamlessly with the resource constrained sensor networks.

Sensors in the network are mostly exposed to open environment as they have to interact with either other sensors or human beings. Physical security of these sensors is always vulnerable and thus poses an unprecedented threat to the overall security of the network. Advances in power analysis and time based attacks enable the malicious entities to perform various hazardous activities.

Wireless channels are still considered unreliable and the same is the case with wireless sensor networks, which may contain a very large number of nodes and sinks, thus giving rise to concerns about the validity of the communications in the network. Trust models for the nodes have to be developed to make sure that all the nodes taking part in the communications are trustworthy.

All these unique features of wireless sensor networks changes the way we look at their security. These networks face different kinds of threats from those of computer, wired, network or even the high-bandwidth wireless models. Thus, these intimidations are coped in distinctive manners.

This chapter will be beneficial in equipping the readers with the basic concepts of security and WSN security. Readers will be able to realize the strengths and weaknesses of WSN with respect to security. Some of the famous and latest attacks and their countermeasures will help in better understanding of the threats and our capabilities to cope with them. Readers with lesser or no prior knowledge of information security will be able to understand this chapter, because basic concepts needed for better apprehension of security issues will be defined.

We are hopeful that the basics provided in this chapter will help the readers to grasp the fundamental concepts of Wireless Sensor Network Security (WSNS), which will empower them to embark on their journey to further explore this ever-expanding field and to find new problems and their solutions in this interesting research and applications field.

General characteristics of WSN are presents in Section 2 of the chapter. These are the properties of these networks which make them the preferred solution in many applications, though they also present limitations on the viable solutions to the security issues in WSN. These attributes are studied with an emphasis on their importance in the security of WSN.

For reliable and secure communications in WSNs, there are some security qualifications that must be fulfilled. These security requirements are given in Section 3.

Threats in WSN are of diverse natures and kinds. Some of the important threats will be discussed in section 4 of this chapter. Countermeasures to some of the described attacks are presented in section 5.

With growing research work in this field, there are many new results that are benefiting us in making the WSN more resistant to attacks and more efficient in their secure implementations in terms of power and memory. Some of the latest research work and implementations of schemes and algorithms are provided in section 6.

Section 7, the last part of the chapter, concludes our discussion. It provides us with summary of the chapter and also outlines the research domains that can be pursued in the coming future related to WSN.

## 2. GENERAL CHARACTERISTICS OF WSN

Wireless sensor networks are unique in many of their features, which are discussed briefly here. These characteristics make them an attractive choice for many applications, and also present the researchers with distinct security challenges.

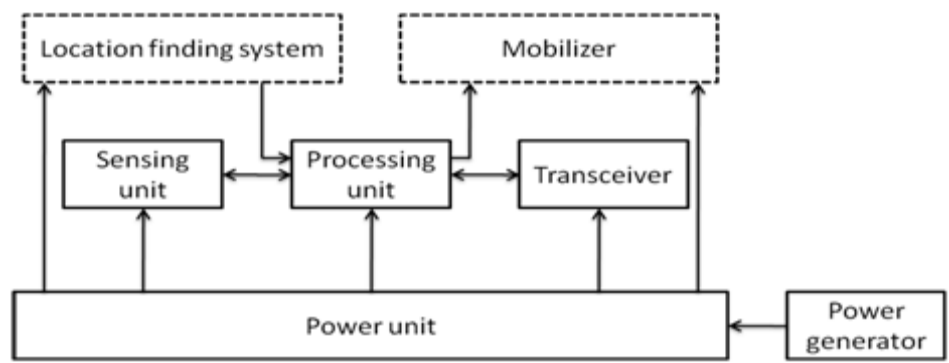

Fig. 1. Sensor node components

## 2.1 Compact Size

As discussed earlier, sensor network may contain hundreds or probably thousand of autonomous nodes. For such a huge network, size does matter. Sensors are kept small, which also limits the components on the main chip-board of the sensor and only the most crucial parts are installed on it.

Small sizes of sensors may be considered as a positive attribute, as sensors can be deployed so that they are not visible.

## 2.2 Physical Security

Sensors usually get information about the environment and perform their designated operations. They have to interact with exposed surroundings which pose hazards to the physical protection of the sensors.

## 2.3 Power

Sensors in WSN contain non-renewable power resources thus causing an energy starved wireless network. Sensors cannot be recharged because of the volume and distribution of the network, which makes recharging of the nodes a laborious and expensive task. Power limitations in WSN are considered the major constraint to the performance of the network. As all the nodes do local processing, they are always in need of power. Thus, the inclusion of security features like encryption, decryption, authentication etc comes at the price of decrease in the overall performance of the nodes because of the energy consumed during these cryptographic algorithms and schemes.

Security is vital for WSN, so there is always some compromise to make between the secure communication and allocation of energy resources for implementing cryptographic schemes.

## 2.3 Memory Space

Sensors have small memory space, which accounts for its low cost and power consumption. Memory is a precious asset for any sensor, thus keeping the size of the security algorithm source code small. Sizes of the keys that need to be stored are also kept at a minimum length because of scarcity of memory storage. Following table lists some of known sensor nodes and their memory spaces.

Table 1. Sensor nodes and their memory spaces

| Sensor Node | Microcontroller | Program and data memory | External memory |
|---|---|---|---|
| IMote 2.0 | Marvell PXA271 | 32 MB SRAM | 32 MB Flash |
| Mica2 | ATMEGA 128L | 4K RAM | 128k Flash |
| TelosB | TI MSP430 | 10k RAM | 48k Flash |
| Ubimote2 | TI's MSP430F2618 | 8k RAM | 116k Flash |

## 2.4 Bandwidth

WSN is a low bandwidth network and as compared to other wireless networks, the quantity of data transmitted and received by the nodes is very low. This helps the nodes in saving the crucial power for other functions. As an estimate, each bit transmitted consumes as much power as executing 800-1000 instructions. This is one of the reasons why cryptographic schemes with large key sizes (i.e. public key cryptography) are not preferred for these sensor networks.

## 2.5 Unreliable Communications

Like all other wireless communications, channels in the WSN are subject to unpredictable environmental conditions, state of channels, interference and many other factors that usually deteriorate the quality of service of the wireless links and induce errors in the information being transmitted.

Error correcting codes, MAC and cyclic redundancy check (CRC) are sometimes used to cope with these problems. They are widely being used in wireless links to ensure better service at the expense of extra bits added to the original messages.

## 3. SECURITY REQUIREMENTS IN WSN

WSN is a wireless network composed of sensors. Due to the attributes of being a network and utilizing wireless communications, the security demands for WSN are unique. Security requirements in WSN to ensure trustworthy and secure connections and communications are a combination of the specifications for computer network and wireless communication security. WSN has its own distinct features, as discussed in section 3, which make these networks unique. Their anomalous character is due to their large volume, pattern of distribution and resource restrictions. All these aspects give rise to some particular security necessities. We will discuss some of basic security specifications for WSN.

## 3.1 Data Confidentiality

Data is communicated between the sender and the recipient, sometimes being routed through many nodes. This data may also be kept in memory for further processing. This data can be sensitive enough to be known only by the sender and the recipient. Sometimes, the adversary can access this information by eavesdropping between wireless links, gaining admission to the storage or by other attacks. Data confidentiality means that the data can only be accessed, and thus utilized, by only those entities that are authorized for this purpose.

If any data is lost by negligence and weak security measures, it can lead to identity thefts, loss in business, privacy breaching and many other malicious activities. This makes data or message confidentiality the most important feature of any security protocol.

In WSN, data confidentiality can be observed by making sure that

**i)** Sensor network should not leak any data to other networks in vicinity, thus retaining the message completely within the network.

**ii)** Data is sometimes routed through many nodes before reaching the destination node. This causes a rise in need for secure communication channels between different nodes and also between nodes and base stations.

**iii)** Encryption is one of the most commonly used procedures to provide confidentiality of data. Critical information such as keys and user identities should be encrypted before transmission. Sensitive information can be characterized from the kind and type of protocol being used i.e. symmetric or asymmetric cryptography, mutual authentication, identity or nonce based encryption.

**iv)** Steps can also be taken towards encrypting the sensitive data before storing them in memory. This is particularly important if the nodes are exposed to user interaction, or in military applications.

Mostly symmetric cryptography or stream ciphers are used for encryption and decryption in WSN, due to the high storage and computational costs associated with the public key cryptography. TinySec is a link layer security protocol that makes use of RC5 and Skipjack block ciphers in cipher block chaining (CBC) mode of operation. LLSP uses AES in CBC mode. LiSP utilizes stream cipher for providing encryption.

## 3.2 Data Integrity

Provision of data confidentiality stops the leakage of data, but it is not helpful against insertion of data in the original message by adversary. Integrity of data needs to be assured in sensor networks, which solidifies that the received data has not been altered or tampered with and that new data has not been added to the original contents of the packet. Environmental conditions and channel's quality of service can also change the primitive message.

Data integrity can be provided by Message Authentication Code (MAC). For this purpose, both sender and receiver share a secret key. Sender computes the MAC using this key and contents of message, and transmits the message along with the MAC to the receiver. The recipient re-calculates MAC by using the shared secret key and message. Absence of irregularity in composition of calculated MAC establishes integrity in the received message.

## 3.3 Data Authentication

Authentication is used in sensor networks to block or restrict the activities of the unauthorized nodes. Any disapproved agent can inject redundant information, or temper with the default packets carrying information. It is particularly important in case of decision making chunks of information. Nodes receiving the packets must make sure that the originator of packets is an accredited source. Nodes taking part in the communication must be capable of recognizing and rejecting the information from illegitimate nodes.

Although data or message authentication can be provided by incorporating calculation of MAC, this symmetric procedure is not recommended for multi-party communication.

Symmetric schemes normally use the calculation of MAC at the sender and receiver ends. It is usually done by the same technique as describes in 3.2 (previous part).

Multi-party communications or broadcasting makes use of asymmetric authentication schemes. Data authentication in broadcasting requires strong trust assumptions, thus giving rise to different categories of trust. For authentication purposes, both of the mutual authentication and one-way authentication method can be used based on trust requirements.

In SPINS [20], authors state that if a sender wants to send authentic data to mutually untrusted receivers, symmetric MAC is not secure since any one of the receivers already knows the MAC key and hence could impersonate itself as the original sender of the message. Then it can forge fake messages and send them to other receivers. SPINS constructs authenticated broadcast from symmetric primitives but it establishes asymmetry by the utilization of delayed key disclosure and one-way function key chains.

LEAP [8], on the other hand uses a globally shared symmetric key for broadcast messages to the whole group. As the group key is shared among all the nodes in the network, steps are taken to update this key through rekeying mechanism if any node is compromised. LEAP exercises an efficient approach to get information about any compromised node.

## 3.4 Data Freshness

Some of the messages are critical enough that extra precautions need to be taken to ensure their correction. Confidentiality and Authentication may not be useful when any old message is replayed by any attacker. Data freshness implies that the received messages are recent, and previous messages are not being replayed. Importance of data freshness becomes evident in networks using shared key operations. During the time taken for transmission of shared key in WSN, replay attack can be carried out by adversary.

Data freshness is categorized into two types based on the message ordering; weak and strong freshness. Weak freshness provides only partial message ordering but gives no information related to the delay and latency of the message. Strong freshness, on the other hand gives complete request-response order and the delay estimation. Sensor measurements require weak freshness, while strong freshness is useful for time synchronization within the network.

To accommodate data freshness, nonce which is a randomly generated number or a time dependent counter can be appended to the data. Messages with previous nonce and old counter numbers are rejected. This guarantees acceptance of only recent data, and thus the freshness in data is achieved.

## 3.5 Availability

Introduction of security scheme in WSN comes at the expense of computational storage and energy costs. Security features in the network may be considered as extra feature by some because of the restrictions it can impose on the availability of the data. Insertion of security can cause earlier depletion of energy and storage resources, causing unavailability of data. Similarly, if security of any one node (especially in central point network management) is compromised or any Denial of Service (DoS) attack is launched, data becomes inaccessible.

Availability of data becomes an important security requirement because of the mentioned arguments. Security protocol should consume less energy and storage, which can be achieved by the reuse of code and making sure that there is minimum increase in communication due to the functioning of security protocols.

Processing within the networking and en-route filtering can be used to subsidize the effects of malicious attacks and other issues that may arise because of increase in communication due to utilization of security scheme. There is also a need to avoid central management scheme in sensor networks as they can affect the availability of data due to single point failures. These steps will also make the network robust against attacks.

## 3.6 Self-organization in WSN

As mentioned in previous sections, one of the characteristics of WSN is their composition and distribution. A typical WSN may have hundreds of nodes performing different operations, installed at various locations. Ad-hoc networks are also sensor networks, having the same flexibility and extensibility. These otherwise attractive properties of WSN pose a serious threat to the overall security situation of the network, raising the importance of a self-organized and robust structure of network.

For using public key cryptography based scheme, an efficient design is needed that takes into account all the situations for sharing the key and is capable of trust management amongst different nodes. Keys can be redistributed between the nodes and base stations to provide key management. Schemes can use symmetric cryptography that applies key predistribution methods.

## 3.7 Secure Localization

WSN makes use of geographical based information for identification of nodes, or for accessing whether the sensors belong to the network or not. Some attacks work by analyzing the location of the nodes. Adversary may probe the headers of the packets and protocol layer data for this purpose. This makes the secure localization an important feature that must be catered during our implementation of security protocol.

## 4. ATTACKS ON WSN

Wireless sensor networks are power constraint networks, having limited computational and energy resources. This makes them vulnerable enough to be attacked by any adversary deploying more resources than any individual node or base station, which may not be a tedious task for the attacker. As described earlier, a typical sensor network may be composed of potentially hundreds of nodes which may use broadcast or multicast transmission. This mode of transmission results in a large volume wireless network with many potential receivers of the transmitted information. This makes a number of attacks such as packet alteration or new packet insertion, capturing of node, reply attacks, denial of service and traffic analysis possible to be performed on any sensor network.

WSN can be cooperatively attacked by colluding in which the adversary makes use of illegitimate nodes with the same capabilities as of network nodes. Deployed malicious nodes can work together to take control of any network node, which can be used further to make damages to the network or to amplify the scope of the attack.

The opponent may have highly capable communication links available to carry out any malicious activity, thus making the countermeasure an expensive task. This is a limitation to the security of WSN as we constantly need inexpensive and small devices as nodes in sensor networks.

Deployment of many nodes of WSN in open and harsh environment poses them another major threat. This compromises their physical security, and if the nodes are not temper-resistant, they can be mishandled and tempered with. Attacks on the physical security of the nodes can cause the node to give away the data stored on it, which may enable the attacker to gain access to critical information such as source code, key and other data which may be crucial for security protocol of the entire wireless network. Making these nodes temper resistant may be able to reduce the effects of side-channel attacks and to enhance the physical security of the network devices, but this may not be the feasible solution as the cost per node increases dramatically if we consider such defenses.

WSN are continuously being used in many critical and sensitive applications. WSN are popular thanks to their ability to incorporate in numerous applications in diverse fields. Health care, security, logistics and military applications are some of the areas of deployment of these wireless networks. It is evident that if the capabilities or functionalities of the sensor network are reduced or endangered, it may cause huge losses in terms of money, resources and may even result in human injuries or fatalities.

This section contains basics of some attacks on WSN, and effects of these attacks on the performance of the wireless networks.

## Threat Models

An attacker may have access only to a few nodes which he or she has compromised. Such attacker is classified as mote class attacker. Alternatively an attacker may have access to more powerful devices such as laptops, hence the definition laptop class attacker. Such attackers have powerful CPUs, great battery power, high power radio transmitter and sensitive antennas at their disposal and pose a much larger threat to the network. For example a few nodes can jam a few radio links where as a laptop can jam the entire network.

Finally, attacks launched on a network may be insider or outsider attacks. In outsider attacks the attacker has no special access to the network. In insider attacks however, the attacker is considered to be an authorized participant of the network.

Such attacks are either launched from compromised sensor nodes running malicious code or laptops using stolen data (cryptographic keys & code) from legitimate nodes.

Now some of the major attacks on WSN are presented. Jamming and physical attacks affect the physical layer of the WSN structure. Collision, exhaustion and unfairness attack types belong to the attacks on data link layer of the WSN.

## 4.1 Denial of Service (DoS)

Jamming nodes of networks, sending continuous messaging without following the system communication protocol (link layer protocols) by any node, malicious attacks and environmental condition may cause resource exhaustion and failures of devices in the WSN. This causes degraded system performance and it is not able to function as expected. These are the forms of Denial of Service (DoS) attacks that intend to affect the functionality of WSN.

These attacks are carried out on the physical, link, routing and transport layers of the WSN architecture. Because of resource limitations of WSN, guarding against these attacks become very costly. Researchers put lot of effort to study these attacks and to devise the methods to minimize their impact on the network.

Now we briefly discuss some of the major types of DoS attacks according to the layers whom they affect. Jamming and physical attacks affect the physical layer of WSNs. Neglect and greed, homing, routing information alteration or spoofing, black holes and flooding belong to the type of attacks on network layer of the WSN architecture.

### 4.1.1 Jamming

Nodes in WSN utilize radio frequencies for the transmission of information, as these sensor networks use wireless channels for communications. Jamming is one of the basic yet detrimental attacks that intend to intervene in physical layer of the WSN structure. It is simply the transmission of the radio signals having the same frequencies as being used by the wireless network.

Jamming causes permanent or temporary suspension of message reception and transmission from the jammed node devices. WSN is widely distributed wireless network, which makes complete jamming an unfeasible attempt. Still jamming of a few nodes in WSN can lead to deterioration in effectiveness of many neighboring nodes.

### 4.1.2 Physical Attacks

As mentioned earlier, WSN devices may be deployed in vast geographical areas and in hostile and harsh environments. Moreover sensor nodes are kept cheap and light weight, which limits any effort to make them temper-proof, their ability to withstand harsh climate or conditions and to avoid or regulate any physical or more sophisticated side-channel attacks.

This makes the WSN nodes highly prone to any physical tempering or other attacks performed on its construction. Nodes can be modified to extract key and other important cryptographic parameters that are crucial for working of any security protocol. Similarly adversary can extract source code which eventually provides attacker the information about the network, which can modify the code to get access into the network. Attacker can replace the nodes with the illegitimate and malicious ones, thus compromising the operation of the whole sensor network.

Physical attacks gives the attacker the ability to alter the nodes and thus the network functioning. These attacks are hard to avoid due to the major characteristics of any WSN to be inexpensive and disperse.

## 4.2 Collisions

Collision is a type of link layer jamming, in which the efficiency of the network is reduced by using the fact that continuous transmission of messages can cause collisions in networks. Collisions cause retransmission of the collided messages and if it happens often then the energy resource of a node can be depleted. Another form of this attack can happen when some part packet is altered, which causes MAC mismatch at the receiver. The corrupted packets are transmitted again, increasing the energy and time cost for transmission. Such an attack when prolonged impels the decrease of network fruition.

### 4.3 Exhaustion

This attack drains the power resources of the nodes by causing them to retransmit the message even when there is no collision or late collision. A node can seek access to any channel deliberately and perpetually, forcing the neighboring nodes to respond continuously.

### 4.4 Unfairness

MAC protocols govern the communications in networks by forcing priority schemes for seamless correspondence. It is possible to exploit these protocols thus affecting the precedence schemes, which eventually results in decrease in service.

### 4.5 Neglect and Greed Attack

During communication between any two nodes in WSN, there may be need to route and re-route packets through many nodes. Transmission from source to destination depends on complete and successful routing of the destined packets. Malicious or compromised node in the way can influence multi-hopping in the network, either by dropping some of packets or by routing the packets towards a false node. This attack also disturbs the functioning of the neighboring nodes, which may not be able to receive or transmit messages.

### 4.6 Homing

Cluster head nodes or the base station neighboring nodes are the most important nodes in WSN. In homing attack, the adversary analysis the network traffic to judge the geographic location of cluster heads or base station neighboring nodes. It can then perform some other kind of attacks on these critical nodes, so as to physically disable them or to capture them which in turn can lead to major damages to the network.

### 4.7 Routing Information Alteration (spoofing)

In this attack, routing information is altered and tempered with. This can create new routing paths, or lengthen or shorten existing routing paths thus increasing the end-to-end latency. It repels or attracts traffic decreasing the quality of service. It can also generate false error messages which disable or increase latency for nodes to access the channel.

### 4.8 Black holes

In WSN, it is possible that nodes are not fully aware with the complete topology of the network because of the large volume of the network. If distance-vector-based protocols are used in these sensor networks, they are highly susceptible to the formation of black holes. Malicious nodes can advertise zero-cost routes to other nodes in the networks, which causes more traffic to flow toward these nodes. Malicious node's neighboring nodes compete for unlimited bandwidth, thus causing resource contention and message disruption. If this state continues, the neighboring nodes may as well exhaust causing a hole in the network. These attacks are also known as "sink hole" attacks.

### 4.9 Flooding

An attacker continuously sends connection establishment requests to a node in this type of resource exhaustion attack. Each of such requests makes the node allocate some resources to serve each request. Persist requests by a malicious node may drain the memory and energy resources of the node under attack.

### 4.10 De-synchronization

In this attack, an adversary can fabricate messages containing any control flags or sequence numbers of previous frames, and transmit them to two connected nodes. These fake messages make the nodes realize as if they have lost their synchronization. Nodes retransmit the assumed missed frames, and if the adversary is capable of persistent transmission of forged messages then the resources of the nodes will be soon depleted. Moreover the connected nodes are not able to share any useful information during this attack, as they delve infinitely in synchronization-recovery protocols.

### 4.11 Interrogation

An interrogation attack exploits the two-way request-to-send/clear-to-send (RTS/CTS) handshake that many MAC protocols use to mitigate the hidden-node problem. An attacker can exhaust a node's resources by repeatedly sending RTS messages to elicit CTS responses from a targeted neighbor node.

### 4.12 Sybil Attack

In this interesting attack, a node can take multiple identities which lead to the failure of the redundancy mechanisms of distributed data storage systems in peer-to-peer networks. Sybil attack functions by its property of representing multiple nodes simultaneously. The Sybil attack is capable of damaging other fault tolerant schemes such as dispersity, multi path routing, routing algorithms, data aggregation, voting, fair resource allocation and topology maintenance. This attack also affects the geographical routing protocols, where the malicious node presents several identities to other nodes in the network and thus appears to be in more than one location at a time. Similarly, during the voting process the malicious node can create additional votes thanks to its ability to present several identities at a time. It can strike the routing algorithms by defining many routes through only one node. Resources of a node can be drained by requests from multiple entities which are in fact exhibited by a single malicious node.

### 4.13 Selective Forwarding

A node may drop partial or complete packets hopping through it, thus disturbing the quality of service in WSN. If all packets are dropped, the neighboring nodes become suspicious and may consider it to be malfunctioning thus finding new routes. Malicious node can selectively forward data to avoid suspicion. It can drop some of the data and passes all other to prevent issues that may arise concerning its performance. Malicious nodes may only allow the data transfer from some selective nodes, giving them the space to alter or suppress data from particular nodes.

This kind of attacks becomes very difficult to detect.

### 4.14 Worm holes

Worm holes are formed by malicious nodes working in different parts of the network. In this attack, the attacker receives messages in one section of network over a low-latency link and sends them to another section of the network. These messages are then replayed in the other part of the network thus forming a worm hole in the present structure of the information flow in network. The impression can be detrimental if the adversary finds its presence near the base stations, giving the distant nodes the realization that they are in the vicinity of the base stations. Multi-hop nodes get the notion through wormholes that they are only one or two nodes away from the base station. Traffic flows to the low-latency route that the adversary provides to these distant nodes. This may cause congestion and further retransmissions of the packets by the legitimate nodes, dissipating their energy.

This attack when used in conjunction with Sybil and selective forwarding attacks becomes difficult to distinguish and evade.

### 4.15 Hello Flood Attacks

At the start of communication, node has to announce itself to the network by broadcasting hello message to their neighboring nodes. It also validates that the node sending hello message is in the vicinity. Adversary can exploit this feature by using a high-powered wireless link. It can assure every node in the network that he is their neighbor, thus starting communication with nodes. As obvious, by using this attack security of the information is compromised as the attacker gains access to the information flow in the network. If some puzzle scheme is used by the nodes to provide access to any node requesting for connection, then a variant of this attack can also be applied.

Adversary should possess enough resources to manage this attack, and should be able to provide high quality routing path to other nodes in network. Traffic will find this path attractive enough to send packets through it, creating data congestion and disturbing the hierarchy of the data flow in network.

## 4.16 Acknowledgement Spoofing

Acknowledgments play a vital role in determining the quality of service at any links and establishing further connections based on the this information. Adversary can alter acknowledgements to present to any transmitting node that any weak link is strong enough for reliable communication.

The packets that are sent on this link are partially or completely dropped, thus decreasing the overall attainment of the WSN.

## 4.17 Node Replication Attack

Sensor nodes have IDs as their identity (and indices of their location in geographical routing algorithms) in the WSN. An adversary can add new node to the sensor network by copying the ID of an already existing node and assigning it to the malicious node. This ensures presence of the adversary in the network allowing the malicious entity to induce destructive affects to the sensor network.

By using the replicated node, packets arriving through it can be dropped, misrouted or altered. This results in incorrect contents of information packet, loss of connection, data loss and high end-to-end latency. Adversary can gain access to the critical information (cryptographic key, source code or other security parameters) by practicing this attack, which brings about security implication of the whole sensor network.

Replicated nodes at specific location can be used to carry out coordinated attack to influence particular nodes or sections of the network.

## 5. COUNTERMEASURES TO ATTACKS ON WSN

## 5.1 Denial of Services (DoS)

### 5.1.1 Jamming

Jamming and its countermeasures depend on the resources of both the sensor nodes and that of the device used by attacker. One of the most obvious solutions to avoid jamming is spread spectrum, or code spreading as used in mobile communication. In these methods, several frequencies are utilized for transmission. Both of these spreading techniques are affective against jamming, as the simple jammer is usually not capable to jam wide band of frequencies or switch to the exact frequencies as being used in frequency hopping or spread spectrum. Implementing these procedures in hardware requires more space, and increases the overall complexity and cost of the device. Sensor devices are kept inexpensive and compact in size, which limits the prospect of deploying these methods in practice.

Jamming attacks can be characterized by high background noises which can be detected and reported by the neighboring nodes. If the jammed part of the WSN is identified, then a deviation in routing paths can help in avoiding this attack.

If jamming attack is found by the network, the sensor nodes under attack can be put to sleep for a long time. Low duty cycle can be applied to consume less power. This enables the nodes to conserve their already limited energy resources, which gives them opportunity to try to connect to WSN once the attack is over. Attacked sensor node can also send a high power message reporting the attack to neighboring nodes or base stations during the attack, if the attacker employs stuttering or interruptive jamming.  Another efficient yet costly solution is the alternative use of optical or infra-red communications for sensor devices under jamming attack, but these modes are distance restricted and quite expensive.

### 5.1.2 Physical Attacks

Adversary can exploit physical weakness of motes to access crucial data stored on it, and is also capable of damaging or replicating the nodes. Steps that one must to ensure the physical safety of sensor nodes in WSN are based on the desired level of security. One cannot fully guarantee complete protection of hundreds or thousands of nodes, which are typically dispersed over large distance to form WSNs.

Nodes in hostile environments can be made temper-proof so that security of these motes is not compromised over cost. Camouflaging and hiding sensor nodes are other countermeasures against physical attacks.

Motes which handle critical data can use any erasure procedure which makes them remove any critical information i.e. cryptographic keys or codes, when they are tempered with.

### 5.2 Collisions

Altered packets of information can increase latency in networks, and results in dropping and discarding of packets once they are found corrupt thus degrading the service of the network. Collision detection and avoidance schemes can be employed to avert such situations. Cyclic redundancy check (CRC) of the messages can be computed on the transmitter and receiver ends to ascertain the integrity of the message. Similarly, error correcting codes can also be used for avoiding and corruption by outsider to the messages. Such codes, with high error correcting capabilities, come at the expense of extra bits that must be appended with the original message. This poses a limitation to the effectiveness of these codes as the malicious agents may be able to inject more errors in the message than the capabilities of the correcting codes. Cooperation between the communicating nodes can also avoid the corruption of the transmitted packets.

### 5.3 Exhaustion

Exhaustion of the power of the sensor due to retransmissions even though they are caused by late collisions, can be handled by use of time division multiplexing (TDM). TDM provides each sensor with a time slot to send its data which avoids collisions. This solves the infinite deference problem, which is caused by continuous retransmissions by nodes.

Allowing limited number of requests to access network at a time can also help in getting rid of collisions. Such a limitation is implemented by exercise of MAC admission control rate, which allows only specific number of requests to access the network.

### 5.4 Unfairness

Adversary exploits the cooperative MAC priority scheme by making sensors to miss their transmission deadlines. This attack affects the real-time users to a large extent. Use of small packets avoids this attack as each sensor node seizes the channel only for short time.

### 5.5 Neglect and Greed Attack

Due to partial drop of packets and unpredictable behavior of malicious node in this attack, it is not possible to detect this type of attack. The best step to avoid damage by neglect or greed of malicious sensor node is to define alternative routing paths. Another proposed solution is to use redundant messages that reduce the impairment by malicious node.

## 5.6 Homing

Adversary learns about the important nodes by analyzing the headers and contents of the messages flowing in the network. Encrypting the header and contents of message makes the task of adversary more difficult. Source and destination of the intercepted messages becomes discreet by using cryptography.

## 5.7 Routing Information Alteration (spoofing)

Routing information included in the packets are altered or spoofed to divert the flow of traffic to the intended destinations. Node addresses can be changed and adversary can control the flow of traffic, which makes it possible for it to attack any particular node.

Packets construction can be made secure by using CRC or MAC schemes, which makes the detection of tempered packets easy. Similarly, link layer authentication also helps to avoid this attack. Only authorized nodes are allowed to take part in exchange of information.

Similarly, interrogation attacks can be handled by the use of authentication and antireplay protection schemes.

## 5.8 Black holes

To counter the formation of black holes, similar steps to that of routing alteration (also termed as "misdirection" in some texts) are taken as this attack also functions by changes in the routing information of the traffic.

Requests for exchange of data should come only from authorized sensors, and an efficient authentication scheme must be deployed to ensure this. WSN can use public key cryptography to sign and verify the routing information and updates. Public key cryptography is quite costly and requires large overhead which makes its utilization for this purpose very difficult. Efficient certification and threshold based cryptography based schemes are advised to be used for authentication and trust management in WSN.

Neighboring nodes can monitor the activities of the node, and can analyze its behavior by sending dummy packets and checking whether it reaches its destination. Geography based probing do not require all nodes in the network to participate in monitoring activities. Physical topology of the network is analyzed by sending probe to detect any black holes and damaged regions.

## 5.9 Flooding

Flooding cause the allocation of resources to the requesting clients and limits the effectiveness of already resource starved sensor node. One method to void this attack is to limit the number of connections. This method has disadvantage of restraining the approval of connection to legitimate nodes at times.

Clients who wish to be connected can be presented puzzles to solve, to show their commitment. Adversary needs to allocate more resources to carry out this attack. Puzzle scheme takes more energy resources than usual of the sensors by it also makes the flooding attacks more studious for the attacker.

Legal nodes need to put more resources to establish connection, which comes as a drawback of this procedure.

## 5.10 De-synchronization

Adversary forges the control fields and the transport layer header to cause retransmissions and eventually lose of synchronization between communicating nodes. Authenticating the critical parts for transportation of the packets provides counter to this kind of assault on motes.

The receiver end detects any fake messages and is able to ignore the instructions carried out by them.

## 5.11 Sybil Attack

Insider node cannot be prevented from launching this attack, but its activities can be restricted. In order to prevent an insider from communicating within the network and establishing shared keys with every node in the network, the base station limits the number of neighbors any sensor can establish connection with. If any node tries to exceed this limit, it results in occurrence of error. By using this scheme, a node when compromised, is limited to communication with only a limited number of nodes which tend to be in its vicinity.

Moreover identities of the nodes which request to establish connections are verified. Each node shares its unique key with the base station. Neighboring nodes exchange information between themselves using the shared key to verify the communication. Compromised node is able to communicate only with its neighbors, thus restraining the affect of this attack.

## 5.12 Selective Forwarding

Like route alteration attack, the step to eradicate or avoid this attack is the use of multipath routing. This measure ensures that the destination finally gets the message sent towards it, through some disjoint path of that of malicious node.

Regular monitoring of the network enables the WSN to track suspicious behavior by any node. Source routing that uses the geographical monitoring of the network can also be used as a prevention measure to this type of attack.

Similar preventions and counter-measures can be applied to other attacks on WSN, as they are also variants of the described attacks.

## 6. LATEST RESEARCH AND IMPLEMENTATIONS

## 6.1 User Authentication

Authentication is one of the foremost security features, and hence it finds its applications in WSN security at different levels. It may include authentication of client nodes to authorize the access to channel or exchange of information, Or it may be in form of signing and verifying messages so as to ensure that the contents of the received messages are intact, which saves network from many attacks.

Now we will briefly mention some of the important works that have been done for authentication process for WSN.

Jaing et al. [22] presents a distributed user authentication scheme in wireless sensor networks. This scheme uses self-certified keys cryptosystem (SCK). They make use of Elliptic Curve Cryptography (ECC) to establish pair-wise keys in their user authentication scheme. This user authentication scheme provides less computational and communication overhead.

Taojun Wu et al. [23] propose a group-based peer authentication scheme for real-time sensor applications. Authenticity and integrity of messages received by base station are crucial in final tracking results. They designed a security component MultiMAC, which uses SkipJack implementation in TinySec as symmetric cipher. Each sensor node stores a different set of keys in its memory, pre-defined by a key mapping scheme. Multiple message authentication code (MAC)s of every message are calculated in SkipJack, using the key set assigned to the sensor node. The receiver authenticates the message by recalculating MACs using its shared keys, thus providing authenticity of received message.

Broadcast authentication limits the number of clients requesting to establish connection by giving access to authorized and trusted nodes, and thus proves to be an important security service. In WSN, digital signatures (sign and verify) and μTESLA-based methods provide broadcast authentication. Both of these techniques can be exploited by the adversary, which results in increase of cost for sensors. Signatures are too expensive to be applied for every connection request and packet forwarding can be used on μTESLA technique, thus making these two methods a weak choice. P. Ning et al. [24] suggests the use of message-specific puzzle scheme to ensure broadcast authentication, which proves to be much better in terms of cost and effectiveness.

In [25], a scheme for cooperative distributed public key authentication scheme that does not require any cryptographic overhead is presented. Each node stores a few number of hashed keys for other nodes. When a public key authentication is needed, the nodes who store this key help in authenticating it in a distributed and cooperative manner.

K. Han et al. [26] in their paper in "Sensors" propose an untraceable node authentication and key exchange protocol. The protocol adds light overhead which intends to increase the lifetime of the sensors. The protocol insures untracebility of the nodes, and works well in dynamic environments.

## 6.2 Key Establishment

Key establishment among nodes of sensor network is an important security aspect. Key establishment is needed for authentication and encryption processes, which are crucial for securing the network against many attacks. Key management maintains stability between sensor nodes in spite of their low operational efficiency.

Key establishement is performed by using public key protocol like Diffie-Hellman (DH), Elliptic curve DH and by using El-Gamal public key scheme.

Q. Huang et al. [27] presents an authenticated key establishment protocol between a sensor and a security manager in a self-organizing sensor network. This hybrid technique uses symmetric key operations instead of public key protocols to reduce the burden on the resource constrained nodes.

In [28], authors propose efficient hybrid key establishment protocol for sensor network self-organized with equal distribution between sensor nodes. This protocol is applicable to distributed environment without control of base station. The scheme combines elliptic curve Diffie-Hellman key establishment with implicated certificate and symmetric key encryption technology.

Efficient implementations of cryptographic key establishment for WSNs pose a challenge to the limited capability nodes. A light weight implementation of elliptic curve Diffie-Hellman (ECDH) key exchange for ZigBee-compliant sensor nodes is given in [29]. This implementation uses ATmega128 processor running the TinyOS operating system and it perform 192 bit prime field elliptic curve cryptography.

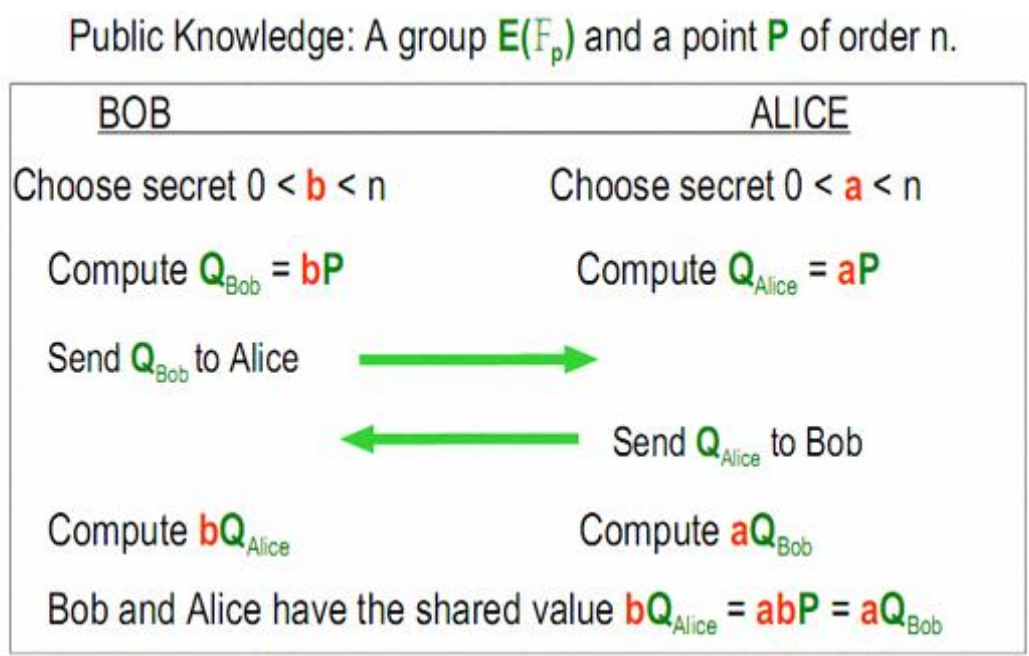

Fig. 2. Elliptic Curve Diffie-Hellman (ECDH)

## 6.3 Trust Management

Trust between the cooperating entities is an important issue in any networked environment. Trust can solve some problems beyond the capability of traditional cryptographic security. It can be used to judge the quality of service being provided by any sensor, which can further help in deciding about provision of access control to that node. In simple networked systems, where security was not deemed necessary, it was assumed that all the parties participating in the communication in the network are trusted ones. But this is not applicable to modern network systems and same is true for wireless sensor networks. We need good trust model within the network to be able to establish connections, exchange keys and information.

M. Momani et al. [30] presents a trust model based on the observed difference in monitoring events and reporting data. This model takes sensor reliability as a component of trust.

H. Chen [31] proposes a task-based trust management framework for WSNs, in which nodes maintain reputation for other nodes of several different tasks and use it to evaluate their trustworthiness. The sensor node maintains a trust rating for different tasks while cooperating with other nodes. The node considers this trust rating to decide its priority to cooperative with nodes with different operations and tasks. A watchdog technique observes the behavior in different task of these nodes and broadcast their trust ratings.

## 6.4 Implementations

Implementations of different cryptographic protocols are widely discussed and researched, due to the difference in their strength, key sizes and application abilities. Some of the latest research in this regard is described here.

R. Roman and C. Alcaraz [32] discusses the possibility of using public key infrastructure in wireless sensor networks, as earlier public key systems were considered too expensive. The authors state that this notion has been partially changed due to development of new hardware and software prototypes based on Elliptic Curve Cryptography (ECC) and other PKC primitives. They point out the possibility to incorporate public key infrastructure such as digital signatures, in the near future.

Hardware implementation of public key cryptosystems is given in [33]. The authors implement 1024-bit RSA and 160-bit ECC public key cryptosystems on Berkeley Motes. They achieve execution times of 0.79 secs for RSA public key operation and 21.5 secs for private operation, and 1.3 secs for ECC signature generation and 2.8 secs for verification. They also implement ECC on Telos B motes with signature time 1.60 secs and a verification time of 3.30 secs.

In ECC, scalar multiplication takes most of the execution time. It has been estimated that nearly 80% of the time is taken by scalar multiplication step. Authors in [34] suggest that there is a room to reduce the key calculation time to meet the potential applications, in particular for wireless sensor networks (WSN) by reducing the time needed for multiplications. They proposed that the positive integer in point multiplication may be re-coded with one's complement subtraction to reduce the computational cost.

A. Liu and P. Ning [35] present the design, implementation scheme, and evaluation of TinyECC, which is a configurable library for implementation of ECC in wireless sensor networks. TinyECC provides a readymade, publicly available software package for ECC-based public key structures. Different optimization steps are included in TinyECC giving the developers the capability to utilize it on different platforms efficiently.

Author in [36] states that most of the public-key cryptographic implemented on small devices are in conjunction with special purpose cryptographic hardware. Accelerators for many crypto functions are used along with small processors.

However in [37], authors implemented ECC without use of any special hardware. With the help of their new algorithm that reduces memory accesses, they achieved 160-bit ECC point multiplication on an Atmel ATmega128 at 8MHz at 0.81 secs. This is the best known execution time for such an operation without using specialized cryptographic hardware.

Software and hardware co-design of ECC $\{GF(2^{191})\}$ is implemented in [16] using Dalton 8051 and special hardware. The hardware consists of an elliptic curve acceleration unit (ECAU) and an interface with direct memory access (DMA) to enable fast data transfer between the ECAU and the external RAM (XRAM) attached to the 8051 microcontroller.

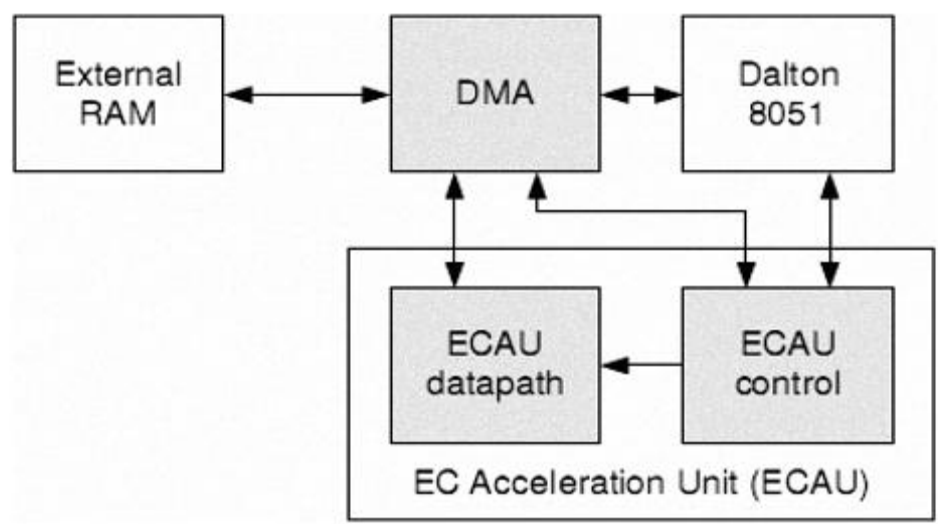

Fig. 3. System block diagram for Software/Hardware co-design of ECC

The special hardware and software combination enables the authors to perform the full scalar multiplication over the field $GF(2^{191})$ in about 118 msecs, assuming that the Dalton 8051 is clocked with frequency of 12 MHz.

Author in [36] shows that ECC can be executed at 63.4 msecs, by using TMS54xx type digital signal processors (DSP). With the decrease in the prices of DSP chips and their compactness, it is safe to think that these processors can be used in WSN sensors in near future.

## 7. CONCLUSION

This chapter serves as a text for researchers especially the beginners, and enables them to get an overview of this ever increasing area of research, wireless sensor networks. This chapter gives a brief yet extensive insight into intriguing world of sensors. Chapter contains many topics of interest, and many more can be found by investigating more deep into this research field.

Chapter has been divided into different sections, describing different aspects of WSN. Basic characteristics of WSN are discusses to give the readers an outline of WSN, which helps in understanding the attacks on WSN and their countermeasures. Some of the major attacks on WSN are given, along with their preventive and counter steps.

The challenges to the field of WSN are unique, and so are their security designs. In time to come, we must be ready to accept many more unique designs of WSN, more sophisticated attacks and their preventions.

**Keywords: wireless sensor network, security, threats, attacks, countermeasures, implementations, survey**

## REFERENCES

1. I. F. Akyildiz, W. Su, Y. Sankarasubramaniam, and E. Cayirci. A survey on sensor networks. IEEE Communications Magazine, 40(8):102–114, August 2002.

2. Mona Sharifnejad, Mohsen Shari, Mansoureh Ghiasabadi and Sareh Beheshti, A Survey on Wireless Sensor Networks Security, SETIT 2007.

3. Hemanta Kumar Kalita and Avijit Kar, Wireless Sensor Network Security Analysis, International Journal of Next-Generation Networks (IJNGN),Vol.1, No.1, December 2009.

4. John Paul Walters, Zhengqiang Liang, Weisong Shi, and Vipin Chaudhary. Wireless Sensor Network Security: A Survey. Security in Distributed, Grid, and Pervasive Computing. Yang Xiao,(Eds.) 2006.

5. T. Aura, P. Nikander, and J. Leiwo. Dos-resistant authentication with client puzzles. In Revised Papers from the 8th International Workshop on Security Protocols, pages 170–177. Springer-Verlag, 2001.

6. R. Anderson and M. Kuhn. Low cost attacks on tamper resistant devices. In IWSP: International Workshop on Security Protocols, LNCS, 1997.

7. D. Boyle and T. Newe,"Securing Wireless Sensor Networks: Security Architectures", Journal of Networks, 2008.

8. X. Du and H. Chen. Security in Wireless Sensor Networks. IEEE Wireless Communications, 2008.

9. J. Granjal, R. Silva and J. Silva, Security in Wireless Sensor Networks. CISUC UC, 2008.

10. Z. Liang and W. Shi. PET: A PErsonalized Trust model with reputation and risk evaluation for P2P resource sharing. In Proceedings of the HICSS-38, Hilton Waikoloa Village Big Island, Hawaii, January 2005.

11. P. Albers and O. Camp. Security in ad hoc networks: A general intrusion detection architecture enhancing trust based approaches. In First International Workshop on Wireless Information Systems, 4th International Conference on Enterprise Information Systems, 2002.

12. R. Anderson and M. Kuhn. Tamper resistance - a cautionary note. In The Second USENIX Workshop on Electronic Commerce Proceedings, Oakland, California, 1996.

13. D. Braginsky and D. Estrin. Rumor routing algorthim for sensor networks. In WSNA '02: Proceedings of the 1st ACM international workshop on Wireless sensor networks and applications, pages 22–31, New York, NY, USA, 2002.

14. Z. Liang and W. Shi. Enforcing cooperative resource sharing in untrusted peer-to-peer environment. ACM Journal of Mobile Networks and Applications (MONET), 10(6):771–783, 2005.

15. Z. Liang and W. Shi. Analysis of recommendations on trust inference in the open environment. Technical Report MIST-TR-2005-002, Department of Computer Science, Wayne State University, February 2005.

16. J. Douceur. The sybil attack. In Proc. of the 1st International Workshop on Peer-to-Peer Systems (IPTPS'02), February 2002. http://www.xbow.com/wireless home.aspx, 2006.

17. H. Chan, A. Perrig, and D. Song. Random key predistribution schemes for sensor networks. In Proceedings of the 2003 IEEE Symposium on Security and Privacy, page 197. IEEE Computer Society, 2003.

18. A. Perrig, J. Stankovic, and D. Wagner .Security in wireless sensor networks. Commun. ACM, 47 (6): 53–57, 2004.

19. A. R. Beresford and F. Stajano. Location Privacy in Pervasive Computing. IEEE Pervasive Computing, 2(1):46–55, 2003.

20. A. Perrig, R. Szewczyk ,J.D. Tygar, V. Wen, and D.E. Culler. Spins: security protocols for sensor networks. Wireless Networking, 8(5):521–534, 2002.

21. Y.-C. Hu, A. Perrig, and D. B. Johnson. Wormhole detection in wireless ad hoc networks. Department of Computer Science, Rice University, Tech. Rep. TR01-384, June 2002.

22. Canming Jiang, Bao Li and Haixia Xu. An Efficient Scheme for User Authentication in Wireless Sensor Networks, vol. 1, pp.438-442, 21st International Conference on Advanced Information Networking and Applications Workshops (AINAW'07), 2007.

23. Taojun Wu, Nathan Skirvin, Jan Werner, Brano Kusy. Group-based Peer Authentication for Wireless Sensor Networks. Talk or presentation, 27, April, 2006.

24. Peng Ning, An Liu, and Wenliang Du. 2008. Mitigating DoS attacks against broadcast authentication in wireless sensor networks. ACM Trans. Sen. Netw. 4, 1, Article 1, February 2008.

25. DaeHun Nyang and Abedelaziz Mohaisen. Cooperative Public Key Authentication Protocol in Wireless Sensor Network. Ubiquitous Intelligence and Computing. Lecture Notes in Computer Science, 2006, Volume 4159/2006, 864-873.

26. Han, Kyusuk; Kim, Kwangjo; Shon, Taeshik. 2010. Untraceable Mobile Node Authentication in WSN. Sensors 10, no. 5: 4410-4429.

27. Huang, Q.; Cukier, J.I.; Kobayashi, H.; Liu, B.; Zhang, J., "Fast Authenticated Key Establishment Protocols for Self-Organizing Sensor Networks", International Conference on Wireless Sensor Networks and Applications (WSNA), ISBN: 1-58113-746-8, pp. 141-150, September 2003.

28. Yoon-Su Jeong and Sang-Ho Lee. 2007. Hybrid Key Establishment Protocol Based on ECC for Wireless Sensor Network. In Proceedings of the 4th international conference on Ubiquitous Intelligence and Computing (UIC '07), Jadwiga Indulska, Jianhua Ma, Laurence T. Yang, Theo Ungerer, and Jiannong Cao (Eds.). Springer-Verlag, Berlin, Heidelberg, 1233-1242.

29. Christian Lederer, Roland Mader, Manuel Koschuch, Johann Großschädl, Alexander Szekely, Stefan Tillich, Energy-Efficient Implementation of ECDH Key Exchange for Wireless Sensor Networks. Information Security Theory and Practices --- WISTP 2009, pp. 112–127. September 2009.

30. Mohammad Momani, Subhash Challa and Khalid Aboura. Modelling Trust in Wireless Sensor Networks from the Sensor Reliability Prospective. Innovative Algorithms and Techniques in Automation, Industrial Electronics and Telecommunications, 2007, 317-321.

31. Haiguang Chen. Task-based Trust Management for Wireless Sensor Networks International Journal of Security and Its Applications, Vol. 3, No. 2, April, 2009.

32. Rodrigo Roman and Cristina Alcaraz. Applicability of Public Key Infrastructures in Wireless Sensor Networks. Public Key Infrastructure, Lecture Notes in Computer Science, 2007, Volume 4582/2007, 313-320.

33. Haodong Wang and Qun Li. Efficient Implementation of Public Key Cryptosystems on Mote Sensors. Information and Communications Security. Lecture Notes in Computer Science, 2006, Volume 4307/2006, 519-528.

34. Xu Huang, Pritam Shah, and Dharmendra Sharma. Fast Algorithm in ECC for Wireless Sensor Network. Proceedings of the International MultiConference of Engineers and Computer Scientists 2010 Vol II, IMECS 2010, March 17 - 19, 2010, Hong Kong.

35. An Liu and Peng Ning. 2008. TinyECC: A Configurable Library for Elliptic Curve Cryptography in Wireless Sensor Networks. In Proceedings of the 7th international conference on Information processing in sensor networks (IPSN '08). IEEE Computer Society, Washington, DC, USA, 245-256.

36. Muhammad Yasir Malik. 2010. Efficient implementation of elliptic curve cryptography using low-power digital signal processor. In Proceedings of the 12th International Conference on Advanced Communication Technology (ICACT'10). IEEE Press, Piscataway, NJ, USA, 1464-1468.

37. Nils Gura, Arun Patel, Arvinderpal Wander, Hans Eberle Sheueling, Chang Shantz, Comparing Elliptic Curve Cryptography and RSA on 8-bit CPUs.

    Can be found at: http://www.research.sun.com/projects/crypto

38. Manuel Koschuch, Joachim Lechner, Andreas Weitzer, Johann Grobschadl, Alexander Szekely, Stefan Tillich, and Johannes Wolkerstorfer, Hardware/Software Co-Design of Elliptic Curve Cryptography o an 8051 Microcontroller, CHES2006, LNCS4249, pp. 430-444, 2006.